\documentstyle[preprint,aps]{revtex}

\def\q{\quad}
\def\sg{\sigma}
\def\lrw{\longleftrightarrow}
\newcommand{\bam}{\left( \begin{array}}
\newcommand{\eam}{\end{array} \right)}
\newcommand{\bamq}[4]{\left( \begin{array}{cccc}{#1}&{#2}&{#3}&{#4}\\}

\begin{document}
\draft
\preprint{Lecce University}
\title{Unified Octonionic Representation
of the 10-13 Dimensional Clifford Algebra
}
\author{
Khaled Abdel-Khalek\footnote{Work supported by an
ICSC--World Laboratory scholarship.\\
e-mail : khaled@le.infn.it}
}
\address{ 
Dipartimento di Fisica - Universit\`a di Lecce\\
- Lecce, 73100, Italy -}

\date{Feb. 1997}
\maketitle
\begin{abstract}
We give a one dimensional octonionic representation of the different
Clifford algebra $Cliff(5,5)\sim Cliff(1,9),\ Cliff(6,6)\sim Cliff(2,10)
$ and lastly $Cliff(7,6)\sim Cliff(3,10)$.
\end{abstract}

\widetext
\vspace{2cm}

Since a long time, it has been conjectured that
there exists a possible connection between the different 
members of the ring division algebra (${\cal R}, {\cal C}, {\cal H},
 {\cal O}$) and the critical dimensions of the Green-Schwarz
superstring action 
\cite{kt,bc,ev}. Especially, the octonionic case has 
gained much attention due to its possible relation to the 10 dimensions
physics \cite{sud,ta,mag,oda,fj,st,engl}.
Not just strings, but even extended to p-branes, octonions are usually
related to different 10 , 11 dimensions p-branes 
\cite{duff} and we would expect that the new
M, F, S theories to be no exception.
In this article, we propose
a unified one dimensional octonionic formulation of the different Clifford
algebra : $Cliff(1,9)\sim Cliff(5,5)\ ; \ Cliff(6,6)\sim Cliff(2,10)$
and lastly $Cliff(7,6)\sim Cliff(3,10)$.

The idea is to know how to translate some real $n\times n$, 
${\cal R}(n)$,
 matrices to their corresponding complex and quaternionic matrices
\cite{rot1}, in general, 
which can be extended to the octonionic algebra 
\cite{my1}.

It is well known from a topological point of view that any
${\cal R}^{2n}$ is trivially a ${\cal C}^{n}$ complex 
manifold and any ${\cal R}^{4n}$ is also
a trivial quaternionic manifold
 ${\cal H}^n$, whereas, any ${\cal R}^{8n}$ is again a trivial
${\cal O}^n$ octonionic manifold. And, as any ${\cal R}^n$ is isomorphic
as a vector space to ${\cal R}(n)$ matrices, we would expect 
\begin{eqnarray}
{\cal R}(2n) \rightarrow {\cal C}(n) ; \label{cl0}\\
{\cal R}(4n) \rightarrow {\cal H}(n) ; \label{cl1}\\
{\cal R}(8n) \rightarrow {\cal O}(n) \label{cl2}
.\end{eqnarray}
To prove this structural 
isomorphism\footnote{Actually the isomorphism does not hold
for the octonionic case as it is evident
that matrix algebra is associative whereas 
octonions are not. Nevertheless, we can find
some translation rules between ${\cal O}$ and
$R^8$.} 
, the idea goes as follows
:
For complex variables, one can represent any complex number
z as an element of ${\cal R}^2$

\begin{eqnarray}
z = z_0 + z_1 e_1,
\equiv Z = 
\left( \begin{array}{c}
z_0 \\
z_1
\end{array} \right).
\end{eqnarray}
The action of 1 and $e_1$ induce the following matrix transformations on 
Z
,
\begin{eqnarray}
1.z &=& z.1 = z \equiv Z = \openone Z ,
\end{eqnarray}
while
\begin{eqnarray} 
e_1.z &=& z.e_1 = z_0 e_1 - z_1 
\\ &\equiv& 
E_1 Z   \\
&=&
\left( \begin{array}{cc}
0 & -1 \\
1 & 0 
\end{array}
\right) 
\left(
\begin{array}{c}
z_0 \\
z_1
\end{array} \right) =
\left( \begin{array}{c}
-z_1 \\
z_0
\end{array} \right). 
\end{eqnarray}
Now, we have a problem, these two matrices $\openone$ and $E_1$
are not enough to form a basis for $R(2)$. 
The solution of our dilemma is easy. We should also
take into account 
\begin{eqnarray}
z^* = z_0 - z_1 e_1 \equiv 
Z^* = \left( \begin{array}{c}
z_0 \\
-z_1
\end{array} \right)
\end{eqnarray}
so, we find
\begin{eqnarray}
1^*.z &=& z^* 
\\ &\equiv&
\openone^* Z = Z^*\\
&=& 
 \left( \begin{array} {cc}
1 & 0 \\
0 & -1 
\end{array} \right) 
\left( \begin{array}{c}
z_0 \\
z_1 
\end{array} \right) = 
\left( 
\begin{array}{c}
z_0 \\
-z_1 
\end{array} \right) , \end{eqnarray}
and  \begin{eqnarray} 
e_1.z^* &=& z^*.e_1 =
z_0 e_1 + z_1  = e_1^* z\\
&\equiv& E_1 Z^* = E_1^* Z \\
&=& 
\left( \begin{array}{cc}
0 & -1 \\
1 & 0 
\end{array} \right) 
\left( \begin{array}{c}
z_0 \\
- z_1 
\end{array} \right) =
\left( \begin{array}{cc}
0 & 1 \\
1 & 0 
\end{array} \right) 
\left( \begin{array}{c}
z_0 \\
z_1 
\end{array} \right)
= \left( \begin{array}{c}
z_1 \\
z_0 
\end{array} \right) 
.
\end{eqnarray}
Having, these four matrices $\{\openone,\openone^*,E_1,E_1^*\}$, 
(\ref{cl0}) is proved\footnote{The relation between 
 $\{\openone^*,E_1,E_1^*\}$ and the quaternionic imaginary units
, defined in the next paragraph is the exact reason for
the possible formulation of the 2 dimensions geometry 
in terms of quaternions \cite{wood}. }.

For quaternions, being non commutative, one should differentiate 
between right and left multiplication,
(our quaternionic algebra is given by $e_i. e_j =  - \delta_{ij} + 
\epsilon_{ijk} e_k,
\  \mbox{and}\ i,j,k = 1,2,3)$,
\begin{eqnarray}
q &=& q_0 + q_1 e_1 + q_2 e_2 + q_3 e_3 
\equiv Q = \left( \begin{array}{c}
q_0 \\
q_1 \\
q_2 \\
q_3 \end{array} \right) 
\quad,\end{eqnarray} then
\begin{eqnarray} 
e_1. q &=& q_0 e_1 - q_1 + q_2 e_3 - q_3 e_2  
\\ &\equiv& E_1 Q
\\ &=& 
\left( \begin{array}{cccc} 
0 & $-1$ & 0 & 0 \\
1 & 0 & 0 & 0 \\
0 & 0 & 0 & $-1$ \\
0 & 0 & 1 & 0
\end{array} \right)
\left( \begin{array}{c}
q_0 \\
q_1 \\
q_2 \\
q_3
\end{array} \right) = 
\left( \begin{array}{c}
-q_1 \\
q_0 \\
-q_3 \\
q_2 \end{array} \right) ,\end{eqnarray}
whereas \begin{eqnarray}
( 1|e_1 ). q &=&
q. e_1 = q_0 e_1 - q_1 - q_2 e_3 + q_3 e_2 
\\ 
&\equiv& 1|E_1~Q = Q E_1
\\ &=& 
\left( \begin{array}{cccc}
0 & $-1$ & 0 & 0\\
1 & 0 & 0 & 0\\
0 & 0 & 0 & 1\\
0 & 0 & $-1$ & 0
\end{array} \right)
\left( \begin{array}{c}
q_0 \\
q_1 \\
q_2 \\
q_3
\end{array} \right) = 
\left( \begin{array}{c}
-q_1 \\
q_0 \\
q_3 \\
-q_2 \end{array} \right) , 
\end{eqnarray}
and so on for the different ($e_2, e_3, 1|e_2, 1|e_3$)
which enable us to find any generic $e_i|e_j$ 
\begin{eqnarray}
e_i|e_j.q  = e_i.1|e_j.q = e_i. q .e_j ,
 \end{eqnarray}
then we have  the possible 16 combinations
${\cal H}|{\cal H}$ 
\begin{equation}
\{1,e_1,e_2,e_3,
1|e_1,e_1|e_1,e_2|e_1,e_3|e_1,
1|e_2,e_1|e_2,e_2|e_2,e_3|e_2,
1|e_3,e_1|e_3,e_2|e_3,e_3|e_3\}.
\label{ml1}
\end{equation}
And their corresponding matrices
\begin{equation} 
\{\openone,E_1,E_2,E_3, 
1|E_1,E_1|E_1,E_2|E_1,E_3|E_1,
1|E_2,E_1|E_2,E_2|E_2,E_3|E_2,
1|E_3,E_1|E_3,E_2|E_3,E_3|E_3\}. 
\label{ml2}
\end{equation}
Using the matrices $\{E_1,E_2,E_3\}$, we
have 
\begin{eqnarray}
E_i \times  E_j = -\delta_{ij} \openone + \epsilon_{ijk} E_k ,
\label{lk1}
\end{eqnarray}
they satisfy the same algebra as their corresponding
quaternionic units $\{e_1,e_2,e_3\}$ i.e they are isomorphic. 
Keep in mind this relation in order to compare it later with the octonionic
case.

We can  deduce the following  group structure
for our quaternionic operators
\begin{itemize}
\item Left $su(2)_L$ 
\begin{eqnarray}
e_i . e_j         &=& -\delta_{ij} + \epsilon_{i j k} e_j , \label{p1} 
\\su(2)_L &\sim& ~\{e_1,e_2,e_3\} . 
\end{eqnarray}
\item Right $su(2)_R$
\begin{eqnarray}
1|e_i . 1|e_j     &=&  1|(e_j.e_i) = -\delta_{ij} + 
\epsilon_{j i k} 1|e_k , \label{p2} 
\\su(2)_R &\sim& ~\{1|e_1,1|e_2,1|e_3\} . 
\end{eqnarray}
This rule can be also explicitly derived using $\{1|E_1,1|E_2,1|E_3\}$ 
.
\item $so(4) \sim su(2)_L \times su(2)_R$, which can
be proved using (\ref{p1}) and (\ref{p2}) and
\begin{eqnarray}
e_i . 1|e_j       &=& 1|e_j . e_i = e_i|e_j 
,
\quad \mbox{i.e} \quad [ e_i ,~1|e_j ] = 0 , \label{kl1}\\
so(4) &\sim& ~\{e_1,e_2,e_3,1|e_1,1|e_2,1|e_3\} . 
\end{eqnarray}
A weak form of (\ref{kl1}), as we will see later, holds for octonionis.
\item $spin(2,3)$ - and its
subgroups - which can be proved by a Clifford Algebra 
\footnote{By explicit
calculation one can show that the gamma matrices given in the next equation
are nothing but the famous Dirac representation up to a minus sign, 
namely,
$-\gamma_0,-\gamma_1,-i\gamma_2,-\gamma_3,-\gamma_5$.} construction
\begin{eqnarray}
&~&\gamma_1 = e_3 ,~~ 
\gamma_2 = e_2 ,~~
\gamma_3 = e_1|e_1 ,~~
\gamma_4 = e_1|e_2 ,~~
\gamma_5 = e_1|e_3 , \label{gamma5}\\
&~&\{\gamma_\alpha , \gamma_\beta \} = 2 diag(-,-,+,+,+) .
\label{clf1}
\end{eqnarray}
By explicit calculation, one finds (in the basis given above)
\begin{eqnarray}
spin(2,3) &\sim&
~\{[ \gamma_\alpha , \gamma_\beta ] \} \quad \quad
\alpha , \beta = 1..5 \quad , \\
&\sim&~\{e_1,1|e_1,1|e_2,1|e_3,
e_2|e_1,e_3|e_1,e_2|e_2,e_3|e_2,e_2|e_3,e_3|e_3\} .
\end{eqnarray}
Actually, the main reason for this construction is
the following relation 
\begin{eqnarray}
e_i e_j ~1|e_k + e_j e_i ~1|e_k = 0 \label{tl1} ,
\end{eqnarray}
this construction is well known since
a long time and used by Synge \cite{Synge} to give a quaternionic formulation
of special relativity ($so(1,3)$) but we don't know who
was the first to derive it (most probable is conway but the
reference is too old and rare to find).
\item  
Also at the matrix level the full set ${\cal H}|{\cal H}$ closes an algebra,
then using the above equations and defining
\begin{eqnarray}
1|e_i . e_j|e_k   &=& \epsilon_{kil} e_j|e_l ,\label{il3}\\
e_i . e_j|e_k     &=& e_j|e_k . e_i = \epsilon_{i j l} e_l|e_k ,\label{il4}\\
e_i|e_j . e_m|e_n &=& \epsilon_{i m l} \epsilon_{n j p} e_l|e_p \label{il5}. 
\end{eqnarray}
By explicit calculations, we found that it is impossible
to construct a sixth $\gamma$ from this set, ${\cal H}|{\cal H}$,
so it is not isomorphic to any $so(n,m)$ algebra!
\item 
Adding the identity to ${\cal H}|{\cal H}$, we used Mathematica
to prove that these 16 matrices are linearly independent
so they can form a basis for any $R(4)$ as we claimed 
in (\ref{cl1}). 
\end{itemize}

A big difference between octonions and quaternions is the following
: All the last equations can be reproduced by matrices {\bf exactly} by
replacing $e \longrightarrow E$ i.e
there is an isomorphism between
(\ref{ml1}) and (\ref{ml2}). The isomorphism
can be derived explicitly between (\ref{lk1})
 and (\ref{p1}) ,then by deriving the suitable rules at the quaternionic
level (\ref{p2},\ref{kl1},\ref{il3},\ref{il4},\ref{il5}),
 it can be extended to the whole set of left and right actions
as well as their mixing. 
In the octonionic case only the 
Clifford algebraic construction resists and holds.

Moving to octonions, we  use the  
symbols $e_i$ to denote the imaginary octonionic units
where $i,j,k = 1 .. 7$ and $e_i. e_j = -\delta_{ij} + 
\epsilon_{i j k} e_k$
such that $\epsilon_{i j k}$ equals 1 for one of the following 
seven combinations \{(123),(145),(176),(246),(257),(347),(365)\} ,
also, we  use the symbol g to represent a generic octonionic
number, $g_i \in {\cal R}$, and 
its corresponding element over ${\cal R}^8$ is 
denoted by G.
As octonions are non-associative, we meet new problems
\cite{my1,schaf}: 
\begin{itemize}
\item First:
Our left and right matrices are no more isomorphic to the octonionic
algebra, for left action, 
we have
\begin{eqnarray}
[E_i,E_j] = 2 \epsilon_{ijk} E_k - 2 [E_i,1|E_j] ,
\end{eqnarray}
while
\begin{eqnarray}
[e_i,e_j] = 2 \epsilon_{ijk} e_k,
\end{eqnarray}
so the isomorphism at the level of algebra is lost and actually
can never be restored as  matrices
are associative but octonions are not. Moreover
the set $\{E_i\}$ alone does not close an algebra. Include
the right action in our treatment is an obligation not a choice, 
 then, we will
be able to find something useful as we will see.

For right action, the situation is the following
\begin{eqnarray}
[1|E_i,1|E_j] = 2 \epsilon_{jik}1|E_k - 2[E_i,1|E_j] ,
\end{eqnarray}
and we have
\begin{eqnarray}
[1|e_i,1|e_j] = 2 \epsilon_{jik} ~1|e_k .
\end{eqnarray}
\item Second: The anticommutation relations hold  at 
the octonionic and matrix level
\begin{eqnarray}
~\{e_i,e_j\}~=~\{1|e_i,1|e_j\}~= -2 \delta_{ij} ,
\label{clif0}
\end{eqnarray}
and  the same for $E_i$ and $1|E_i$ ,
\begin{eqnarray}
~\{E_i,E_j\}~=~\{1|E_i,1|E_j\}~= -2 \delta_{ij} \openone.
\label{clif1}
\end{eqnarray}
 So a Clifford algebraic construction will be possible.
\item Third: Due to the non-associativity, 
\begin{eqnarray}
(e_1.(e_2 .g)) \neq ((e_1.e_2).g) ,
\end{eqnarray}
we have to introduce left/right
octonionic operators 
($\times$ is the usual matrix multiplication), 
\begin{eqnarray}
e_i(e_j. g = e_i.(g. e_j ) \equiv R_{ij} \times G ,\\
e_i)e_j. g = (e_i. g).e_j \equiv L_{ij}  \times G ,
\end{eqnarray}
which can be constructed from the following sets,
$\{ e_1,\ ...\ , e_7, 1|e_1,\ ... \ , 1|e_7\}$ and
$\{ E_1,\ ...\ , E_7, 1|E_1,\ ... \ , 1|E_7\}$, 
as follows
\begin{eqnarray}
e_i(e_j.g = e_i.1|e_j.g \equiv R_{ij} = E_i
\times 1|E_j \times G ,\\
e_i)e_j.g = 1|e_j.e_i.g \equiv L_{ij} = 1|E_j \times E_i \times G .
\end{eqnarray}
\end{itemize}

The easiest way to construct a Lie algebra from
our left/right octonionic operator is to use a
Clifford algebraic construction.
As it is clear from  (\ref{clif0}), any of the 
set $\{e_i\}$ or $\{1|e_i\}$ gives an octonionic representation
of Cliff(7,0) which can be represented by the matrices
$\{E_i\}$ or $\{1|E_i\}$. 
\begin{itemize}
\item Matrix representation of $so(7)_L$
\begin{eqnarray}
so(7) \sim \{ [E_i,E_j] \} \quad \quad i,j = 1...7
\end{eqnarray}
\item Matrix representation of $so(7)_R$
\begin{eqnarray}
so(7) \sim \{ [1|E_i,1|E_j] \} \quad \quad i,j = 1...7
\end{eqnarray}
\item Matrix representation of $so(8)_L$
\begin{eqnarray}
so(8) \sim S_7\times so(7) \sim 
\{E_i, [E_i,E_j] \} \quad \quad i,j = 1...7
\end{eqnarray}
\item Matrix representation of $so(8)_R$
\begin{eqnarray}
so(8) \sim S_7 \times so(7) \sim
\{ 1|E_i, [1|E_i,1|E_j] \} \quad \quad i,j = 1...7
\end{eqnarray}
where $S_7$ is the Reimannian seven sphere.
\end{itemize}

In summary, whatever
our left/right matrices do not form an isomorphic
representation of our left/right octonionic
operators, they admit an isomorphic Clifford algebra.
Now, trying to have something larger than
Cliff(7,0) like the 
 quaternionic Cliff(2,3) (eqn. \ref{clf1}),
one would try
\begin{eqnarray}
\gamma_0 &\rightarrow& e_2 , \quad
\gamma_1 \rightarrow e_3 ,\quad
\gamma_2 \rightarrow e_4 ,\nonumber \\
\gamma_3 &\rightarrow& e_5 ,\quad
\gamma_4 \rightarrow e_6 ,\quad
\gamma_5 \rightarrow e_7 ,\nonumber \\
\gamma_6 &\rightarrow& e_1(e_1 ,\quad
\gamma_7 \rightarrow e_1(e_2 ,\quad
\gamma_8 \rightarrow e_1(e_3 ,\nonumber \\
\gamma_9 &\rightarrow& e_1(e_4 ,\quad
\gamma_{10} \rightarrow e_1(e_5 ,\quad
\gamma_{11} \rightarrow e_1(e_6 ,\nonumber \\
\gamma_{13} &\rightarrow& e_1(e_7. 
\label{kliff}
\end{eqnarray}

This construction works well for $\gamma_{0..5}$ but fails
elsewhere, for example
\begin{eqnarray}
\{ \gamma_0,\gamma_1 \} g &=&
e_2 ( e_3 ( g_0 e_0 +
g_1 e_1 + g_2 e_2 + g_3 e_3 + g_4 e_4 + g_5 e_5 
+ g_6 e_6 + g_7 e_7 ) )
\nonumber \\ 
&+& 
e_3 ( e_2 ( g_0 e_0 +
g_1 e_1 + g_2 e_2 + g_3 e_3 + g_4 e_4 + g_5 e_5 
+ g_6 e_6 + g_7 e_7 ) )
\nonumber \\
&=& 0.
\end{eqnarray}
whereas
\begin{eqnarray}
\{ \gamma_0 , \gamma_8 
\} &=& 
e_2 ( e_1 ( ( g_0 e_0 +
g_1 e_1 + g_2 e_2 + g_3 e_3 + g_4 e_4 + g_5 e_5 
+ g_6 e_6 + g_7 e_7 ) e_3 ) )
\nonumber \\
&+&
e_1 ( ( e_2  ( g_0 e_0 +
g_1 e_1 + g_2 e_2 + g_3 e_3 + g_4 e_4 + g_5 e_5 
+ g_6 e_6 + g_7 e_7 ) ) e_3 )
\nonumber \\
&\neq& 0. \label{nml1}
\end{eqnarray}
One may give up and say octonions are different from 
quaternions and they are non-associative. But, because of
this reason, we still have more freedom.
By a careful analysis of (\ref{nml1}), it becomes clear
that the reason of the failure is
\begin{eqnarray}
E_i \times ~1|E_j \neq 1|E_j \times E_i
\end{eqnarray} 
But a weaker form holds
\begin{eqnarray}
E_i \times ~1|E_i = 1|E_i \times E_i
\end{eqnarray} 
in complete contrast with (\ref{kl1}). The solution can
be found to get around this problem.
 
Because of the non-associativity, we should give to
left and right action different priorities.
As a matter of fact, this is a very reasonable requirement.
When we transferred from complex numbers to quaternions,
we introduced barred operators in order to overcome the
non-commutativity problem and we defined their consistent
rules, so going to octonions, we should need more rules.

Assuming higher priority to right action i.e
\begin{eqnarray}
e_1(e_2.e_4.g \equiv (e_1.(e_4.(g.e_2))) , \\
e_4.e_1(e_2.g \equiv  (e_4.(e_1.(g.e_2))) .
\end{eqnarray}
then
\begin{eqnarray}
\{\ e_1(e_2\ ,\ e_4 \ \}. g = 0.
\end{eqnarray}
Using these simple rules, we can generalize
 (\ref{gamma5}). 
Using the following identities
\begin{eqnarray}
\{ E_i, E_j \} = -2 \delta_{ij} ,\\
\{ 1|E_i, 1|E_j \} = -2 \delta_{ij} ,\\
E_i \times E_j \times 1|E_k + E_j \times E_i \times 1|E_k 
= 0 ,
\end{eqnarray}
which hold equally well at the octonionic level
\begin{eqnarray}
\{ e_i, e_j \} = -2 \delta_{ij} ,\\
\{ 1|e_i, 1|e_j \} = -2 \delta_{ij} ,\\
e_i . e_j . 1|e_k + e_j . e_i . 1|e_k 
= 0 ,
\end{eqnarray}
in complete analogy with (\ref{tl1}).
Now, we have the possibility to write down
the $Cliff(7,6)$ which are given in (\ref{kliff}).

When any of the $\gamma_{6..13}$'s are translated into matrices,
each one has  two different forms, depends of being
acted from right or left, e.g.
\begin{eqnarray}
\gamma_0 \gamma_9 = 
e_2.e_1(e_4.g \equiv E_2 \times E_1 \times 1|E_4 \times G ,
\label{lp1}\\
\gamma_9 \gamma_0 = 
e_1(e_4.e_2.g \equiv E_1 \times E_2 \times 1|E_4 \times G,
\label{lp2}
\end{eqnarray} 
{\bf they don't have a faithful $8\times 8$ matrix representation}.
To be clear, in (\ref{lp1}), we say that $\gamma_9$ is represented
by the matrix $E_1 \times ~1|E_4$ but in (\ref{lp2}) this 
statement is not valid anymore as $E_2$ is now sandwiched
between the $E_1$ and $1|E_4$.
This is a very important fact and should be always taken into
account. 
When we count the numbers of degrees of freedom
, we have 64 for left action and 64 for right action
, in total 128 real parameters which are enough
to represent our $Cliff(7,6)$. 

Actually, because octonions are non-associative, sometimes, 
we can do
with them what we can not do with matrices in a straightforward
way.

Finally, we want to comment about the possible further
applications and investigations:

1- The Green-Schwarz string action in $D=10$
 depends on a 16-real components Majoranna-Weyl
spinor, the $\kappa$ symmetry removes half of these fermionic 
degrees of freedom leaving the action depends on just
8 real fermionic components i.e one octonion \cite{ced}.
Since, there is {\bf no way} to find 
D=10 dimensions Clifford algebra $8\times 8$ gamma matrices,
this represents an obstacle towards a covariant string formulation.
Our representation is dependent on exactly one octonion i.e 8 real components.
Actually, this was the main motive of this work. Superstring exists
and without any doubt it is our best candidate for the dreamed theory
of every thing, finding its true formulation is highly required.
 Can it be the octonionic string \cite{st}!

2- The unified 10-13 dimensions octonionic 
representation is in agreement with the recent discovery of 13
hidden dimensions in string theory \cite{bar}.
 It would be easier to
work with one octonionic construction
instead of 32 components gamma matrices.

3- What is the real meaning of the different p-branes dualities?
May be nothing but a non-trivial mapping between
their different -- postulated -- infinite dimensional world-volume symmetries.
Or more attractively, different mapping between
different infinite dimensional ring-division superconformal algebra
which may be the real connection between the ring-division algebra and the 
p-brane program. One of the simple formula that 
holds for many p-branes is
\begin{eqnarray}
D - p = 2^n \quad \quad n =  0, 1, 2, 3.
\end{eqnarray}
Does it really mean that any consistent p-brane
should enjoy a superconformal algebra on its transverse
dimensions ?
This can be an amplified form of our old problem, what is the
correct relation
 between the string sheet and the
target space formulation of string theory?

We understand that the approach discussed here 
may not be the best in the market but with our potentially need
of developing and examining the recent string dualities, it seems
worthwhile to try every possible avenue.

\vspace{1cm}

I would like to acknowledge 
P. Rotelli and S. De~Leo as well as the physics department
at Lecce university for their kind hospitality.
Also, I am grateful to Prof. A.~Zichichi 
and the ICSC--World Laboratory for financial
support.

\newpage
\appendix
\section*{}

We introduce the following notation:

\begin{eqnarray}
\{~a, \; b, \; c, \; d~\}_{(1)} ~ &\equiv& ~\bamq{a}{0}{0}{0} 0 & b & 0 & 0\\
0 & 0 & c & 0\\ 0 & 0 & 0 & d \eam    \q , \\ 
\{~a, \; b, \; c, \; d~\}_{(2)} ~ &\equiv& ~\bamq{0}{a}{0}{0} b & 0 & 0 & 0\\
0 & 0 & 0 & c\\ 0 & 0 & d & 0 \eam \q ,\\
\{~a, \; b, \; c, \; d~\}_{(3)} ~ &\equiv& ~\bamq{0}{0}{a}{0} 0 & 0 & 0 & b\\
c & 0 & 0 & 0\\ 0 & d & 0 & 0 \eam    \q , \\
\{~a, \; b, \; c, \; d~\}_{(4)} ~ &\equiv& ~\bamq{0}{0}{0}{a} 0 & 0 & b & 0\\
0 & c & 0 & 0\\ d & 0 & 0 & 0 \eam \q ,
\end{eqnarray}
where $a, \; b, \; c, \; d$ and $0$ represent $2\times 2$ real matrices.

In the following  $\sg_{1}$, $\sg_{2}$, $\sg_{3}$ represent the 
standard Pauli matrices.

\begin{eqnarray}
e_{1}  ~\lrw~\{-i\sg_{2}, -i\sg_{2}, -i\sg_{2},  
i\sg_{2} ~\}_{(1)} \q &,&
1\mid e_{1} ~\lrw~\{-i\sg_{2},  i\sg_{2}, i\sg_{2},  
-i\sg_{2} ~\}_{(1)}\q , \nonumber\\
e_{2}  ~\lrw~\{ -\sg_{3}, \sg_{3}, -1, 1 ~\}_{(2)}\q &,&
1\mid e_{2}~\lrw~\{ -1, 1, 1,  
-1 ~\}_{(2)}\q , \nonumber\\
e_{3} ~\lrw~\{ -\sg_{1}, \sg_{1}, -i\sg_{2},  
-i\sg_{2} ~\}_{(2)}\q &,&
1\mid e_{3} ~\lrw~\{ -i\sg_{2}, -i\sg_{2}, i\sg_{2},  
i\sg_{2} ~\}_{(2)}\q , \nonumber\\
e_{4} ~\lrw~\{ -\sg_{3}, 1, \sg_{3}, -1 ~\}_{(3)}\q &,&
1\mid e_{4} ~\lrw~\{ -1, -1, 1,  
1 ~\}_{(3)}\q , \nonumber\\
e_{5} ~\lrw~\{ -\sg_{1}, i\sg_{2}, \sg_{1},  
i\sg_{2} ~\}_{(3)}\q &,&
1\mid e_{5} ~\lrw~\{ -i\sg_{2}, -i\sg_{2}, 
-i\sg_{2},  
-i\sg_{2} ~\}_{(3)}\q , \nonumber\\
e_{6}~\lrw~\{ -1, -\sg_{3}, \sg_{3}, 1 ~\}_{(4)}\q &,&
1\mid e_{6} ~\lrw~\{ -\sg_{3}, \sg_{3}, -\sg_{3},  
\sg_{3} ~\}_{(4)}\q , \nonumber\\
e_{7} ~\lrw~\{ -i\sg_{2}, -\sg_{1}, \sg_{1},  
-i\sg_{2} ~\}_{(4)}\q &,&
1\mid e_{7} ~\lrw~\{ -\sg_{1}, \sg_{1}, -\sg_{1},  
\sg_{1} ~\}_{(4)}\q .
\end{eqnarray}

\end{document}